\documentclass[runningheads]{llncs}
\usepackage[T1]{fontenc}
\usepackage{amsmath,amssymb}
\usepackage{booktabs}
\usepackage{tabularx}
\usepackage{subcaption}
\usepackage{graphicx}
\usepackage{xcolor}

\begin{document}

\title{Analysis of Public Schools Educational Performance Based on Causal Models and Hierarchical Clustering}

\titlerunning{Causal Models and Hierarchical Clustering}

\author{
Anderson L. de Paula\inst{1,2} \and 
Pedro C. dos Santos\inst{1,2} \and
Renato A. Krohling\inst{1,2}
}

\institute{Labcin - Nature Inspired Computing Lab, Federal University of Esp\'irito Santo, Vit\'oria, Brazil \and
PPGI - Graduate Program in Computer Science, Federal University of Esp\'irito Santo, Vit\'oria, Brazil\\
\email{anderson.vwd@gmail.com},
\email{pcortesdosantos@gmail.com}, 
\email{krohling.renato@gmail.com}
}

\authorrunning{De Paula, Dos Santos and Krohling}

%


\maketitle              

\begin{abstract}
The increasing availability of large-scale educational datasets has expanded the use of quantitative methods for investigating school performance. However, institutional heterogeneity among schools and the structural complexity of educational data pose substantial challenges to traditional statistical modeling approaches. This study investigates the existence of school typologies based on structural, pedagogical, and demographic characteristics, and examines how these typologies relate to performance in the Brazilian Basic Education Assessment System (Saeb). Using data from the Brazilian School Census and Saeb, data preprocessing and normalization procedures are applied followed by hierarchical clustering to identify groups of schools with similar structural profiles. After the identification of these typologies, causal analysis techniques are employed to investigate potential causal relationships between school characteristics and educational outcomes. The results reveal the presence of distinct school profiles and statistically significant differences in average performance among them. The causal analysis provides insights into the structural and contextual factors that may influence educational performance, contributing to a better understanding of the mechanisms associated with school effectiveness.


\end{abstract}
\section{Introduction}
The evaluation of educational performance is one of the main instruments for monitoring the quality of basic education. In Brazil, indicators derived from large-scale assessments, such as the Basic Education Assessment System (SAEB), are widely used to support public policies, institutional diagnoses, and academic research. However, the interpretation of these indicators, when carried out in isolation, it tends to neglect the structural complexity of schools and the multiple dimensions that influence educational outcomes.
    
Schools are not homogeneous units. Differences related to infrastructure, the socioeconomic profile of students, teacher training, and pedagogical organization result in distinct institutional contexts, in which the same performance indicators can express different realities. In this sense, the search for school typologies appears as a relevant analytical strategy to capture underlying structural patterns in educational data \cite{hanushek2003failure}.
    
In this context, data science offers a range of methods capable of identifying latent structures in complex educational datasets, particularly through clustering techniques. However, most existing approaches rely primarily on observable characteristics without explicitly considering the underlying causal mechanisms that shape educational outcomes.
    
This limitation may lead to groupings that are statistically coherent but difficult to interpret from a policy or pedagogical perspective. In particular, clusters derived solely from correlational structures may fail to reflect meaningful differences in school contexts and performance dynamics.
    
To address this gap, this study investigates whether the incorporation of causal variables can improve the identification of school typologies. Specifically, we compare clustering structures obtained under two scenarios: a baseline approach using conventional variables and an alternative approach incorporating variables selected based on causal relevance.
    
The central hypothesis is that causal clustering not only alters the structural organization of groups but also enhances their interpretability and their ability to capture meaningful differences in educational performance. To evaluate this hypothesis, we analyze (i) structural differences between clusterings, (ii) differences in student performance across clusters, and (iii) the internal quality of the resulting partitions.

Research on educational data analysis, causal inference, and hierarchical modeling has evolved along intersecting methodological lines that inform the present study. Early contributions emphasized the importance of modeling the nested structure of educational systems. Raudenbush and Bryk \cite{raudenbush1986} introduced hierarchical linear models as a rigorous framework for analyzing student outcomes while accounting for school-level variation, demonstrating that ignoring multilevel structure leads to biased estimates of school effects. More recent work by García-Jiménez et al. \cite{garcia2022} extends this tradition by applying multilevel models to identify high- and low-efficiency schools in Spain, showing that socioeconomic and cultural factors must be controlled to obtain fair estimates of institutional effectiveness.

Parallel to the development of hierarchical modeling, several studies have explored causal mechanisms underlying student achievement. Prasertcharoensuk et al. \cite{prasert2020} proposed a structural causal model linking transformational leadership, teacher collective efficacy, and professional learning communities to student performance in Thailand, highlighting the interplay between organizational and instructional factors. Rajagukguk et al. \cite{rajagukguk2023} advanced this line of inquiry by applying NOTEARS and Bayesian networks to Indonesian educational data, uncovering causal pathways between socioeconomic variables and academic outcomes. At a more theoretical level, Weinstein and Blei \cite{weinstein2026} introduced Hierarchical Causal Models, a generalization of structural causal models that explicitly incorporates nested data structures. Their work shows that hierarchical organization can enable causal identification even when traditional flat models fail, offering a powerful foundation for causal inference in education.

A complementary body of literature focuses on clustering techniques to uncover latent patterns in student performance. Bowers \cite{bowers2010} demonstrated that hierarchical cluster analysis applied to longitudinal K–12 grade histories can reveal distinct academic trajectories and accurately identify students at risk of dropping out. Nafuri et al. \cite{nafuri2022} applied k-means, BIRCH, and DBSCAN to classify low-income Malaysian students based on behavioral and academic attributes, illustrating the value of unsupervised learning for identifying vulnerable subgroups. Oviedo et al. \cite{oviedo2016} proposed a hierarchical clustering method grounded in probabilistic graphical models, suitable for educational datasets with highly correlated socioeconomic indicators.

More recently, clustering has been integrated with causal inference. Kim et al. \cite{kim2024} expanded the causal k-means framework by introducing hierarchical and density-based causal clustering algorithms capable of identifying subgroups with heterogeneous treatment effects without relying on strong structural assumptions. Their work demonstrates that clustering in counterfactual space can reveal meaningful subpopulations that respond differently to interventions, offering a bridge between unsupervised learning and causal analysis.

Together, these studies highlight the importance of combining hierarchical modeling, causal inference, and clustering to understand complex educational phenomena. They collectively show that student outcomes emerge from multilevel causal processes and that unsupervised methods can reveal hidden structures essential for targeted interventions and policy design.
   
This paper is structured as follows: Section 2 describes the data and indicator construction. Section 3 presents the theoretical background, including causal discovery and validation metrics. Section 4 details the methodology, including the clustering framework under baseline and causal scenarios. Section 5 reports the results, focusing on structural differences, educational performance, and cluster quality. Section 6 discusses the findings, and Section 7 concludes the paper and outlines future work.
     
\section{Data and Construction of Indicators}
\subsection{Databases}
Every year, the National Institute for Educational Studies and Research Anísio Teixeira (INEP) conducts the School Census, which constitutes the main census database for basic education in Brazil. The School Census data collection processes and makes available to school systems a comprehensive set of administrative, pedagogical, and structural information about school units. Since it is not a public database, access to the 2023 School Census data used in this study occurred within the scope of institutional activities of the state public administration, fully respecting the legal and ethical confidentiality criteria established by INEP.
        
The School Census structures its information in various scenarios, such as school, class, enrollment (students), teachers, and administrators. In this study, information related to school units, classes, enrollment, and teachers was used. Thus, the database was composed of attributes distributed across four main items: student characteristics, school information, class data, and school structural characteristics.
        
On the other hand, public data from the Basic Education Assessment System (Saeb), made available by INEP, were used. Unlike the School Census, Saeb aims primarily to measure the educational performance of students through standardized assessments. In this study, Saeb scores aggregated by school and by educational stage were used, covering Elementary School — Initial Years and Final Years — and High School.
        
Considering that a single school can simultaneously offer several stages of education, the Saeb scores were aggregated by school and by stage and then normalized by dividing each score by the standard deviation corresponding to its respective stage. From the normalized scores, a weighted score per school unit was calculated, defined as the average of the scores weighted by the number of enrollments, all segmented by stage.
        
This approach allowed for the comparison of scales across different stages and the construction of a synthetic indicator of school performance, capable of reflecting, in an integrated way, the performance of school considering simultaneously its different educational offerings. Furthermore, it avoided excessive fragmentation of the dataset, which could compromise the statistical robustness of the analyses.

\subsection{Construction of Indicators}
The indicators were structured based on different aspects of the school environment. At the school level, indicators related to the socio- territorial context and infrastructure were developed, such as: differentiated socio-cultural location (indigenous lands, quilombola communities and settlements), availability of electricity, internet access for students, percentage of air-conditioned classrooms, urban or rural location and complexity of the educational offer, understood as the diversity of stages and modalities of education offered by the school (youth and adult education, vocational education, elementary education – initial and final years – and regular or integrated secondary education).
    
\begin{table}[htbp]
\centering
\renewcommand{\arraystretch}{1.4}
\caption{Description of the Dataset Variables}
\label{tab:variaveis_refinada}
\small
\begin{tabularx}{\textwidth}{p{3.8cm} c >{\raggedright\arraybackslash}X}
\toprule
\textbf{Dimensão} & \textbf{N.º Var.} & \textbf{Examples of Variables.} \\
\midrule
        
Identificação (ID) & -- &
\texttt{CO\_ENTIDADE} \\
        
Infraestrutura Básica & 4 &
\texttt{IN\_ENERGIA}, \texttt{IN\_ESGOTO\_REDE\_PUBLICA}, \texttt{IN\_LIXO\_SERVICO\_COLETA}, \texttt{IN\_LOCAL\_FUNC\_PRISIONAL\_SOCIO} \\
        
Recursos e Espaços & 6 &
\texttt{IN\_AUDITORIO}, \texttt{IN\_BIBLIOTECA\_SALA\_LEITURA}, \texttt{IN\_LABORATORIO\_CIENCIAS}, \texttt{IN\_QUADRA\_ESPORTES}, \texttt{QT\_SALAS\_UTILIZADAS}, \texttt{PER\_SALAS\_UTILIZA\_CLIMATIZADAS} \\
        
Tecnologia e Gestão & 4 &
\texttt{IN\_INTERNET\_ALUNOS}, \texttt{QT\_PROF\_COORDENADOR}, \texttt{COMPLEXIDADE\_OFERTA}, \texttt{COORDENACAO\_POR\_ETAPA} \\
        
Fluxo e Matrículas & 8 &
\texttt{QT\_TURMAS}, \texttt{PER\_TURMAS\_REGULAR}, \texttt{PER\_MATRICULAS\_REGULAR}, \texttt{MEDIA\_ALUNOS\_TURMA\_REGULAR}, \texttt{PERC\_TURMAS\_SUPERLOTADAS\_REGULAR}, \texttt{PER\_MATRICULAS\_EJA}, \texttt{PER\_MATRICULAS\_AEE}, \texttt{QT\_MATRICULAS\_TOTAL} \\
        
Perfil Discente & 3 &
\texttt{PER\_PRETO\_PARDO}, \texttt{PER\_SEXO\_FEMININO}, \texttt{PER\_DISTORCAO\_IDADE\_SERIE} \\
        
Corpo Docente & 5 &
\texttt{FORMACAO\_SUPERIOR}, \texttt{SUPERIOR\_INT\_PUBLICA}, \texttt{POS\_GRADUACAO}, \texttt{PROFESSOR\_EFETIVO}, \texttt{IN\_COMPLEMENTACAO\_PEDAGOGICA} \\
        
\bottomrule
\end{tabularx}
        
\vspace{2pt}
\begin{flushleft}
\footnotesize
\textit{Note:} The table presents the 33 analytical variables used in the study, organized by dimension. \texttt{CO\_ENTIDADE} is used exclusively as an identifier. \texttt{IN\_} variables are binary; \texttt{PER\_} indicates proportions; \texttt{QT\_} correspond to absolute quantities.
\end{flushleft}
\end{table}
           
Furthermore, an indicator was developed that links the presence of pedagogical coordination with the complexity of the educational offerings of school. At the classroom level, indicators associated with pedagogical organization and capacity were developed, such as: proportion of overcrowded classrooms, number of regular classes, and average number of students per class. Overcrowding was defined based on an empirical criterion, considering as overcrowded those classes whose number of students exceeds the 0.75 quantile of the distribution corresponding to each stage of education.
    
In the student context, indicators related to student composition and educational trajectory were constructed, including age-grade distortion and distribution by sex and race/color. Additionally, variables related to the distribution of enrollments by educational modality (regular, Youth and Adult Education, and special education – Specialized Educational Services (AEE)) were included. For the teaching staff, indicators related to qualifications and professional affiliation were considered, such as: higher education, postgraduate studies, permanent employment, and participation in pedagogical enrichment programs.
    
At the end of the selection and construction process, a dataset composed of 33 variables was obtained, representing multiple structural, organizational , and pedagogical dimensions of the schools analyzed. These variables were selected with the objective of characterizing each school according to its structural, social, economic, and, to a limited extent, pedagogical profile, since only part of the teachers' information was included. The decision was made not to incorporate school flow, previous Saeb scores, and municipal location in order to avoid the grouping being influenced by geographical or historical factors, focusing the analysis on the current profile of the school unit. All selected variables can be viewed in Table ~\ref{tab:variaveis_refinada}.

\section{Background}
\subsection{Causal Discovery}
\label{subsec:causal-discovery}

Causal discovery aims to infer causal relationships among observed variables directly from data, representing them as a Directed Acyclic Graph (DAG) $\mathcal{G} = (V, E)$, where directed edges encode direct causal effects \cite{pearl2009,spirtes2000}. Unlike associational methods, causal graphs encode asymmetric generative mechanisms and support reasoning about interventions \cite{pearl2009}. The task is inherently challenging: the number of possible DAGs grows super-exponentially with the number of variables, and multiple graphs may encode the same conditional independencies, forming a Markov equivalence class. Existing algorithms address this through distinct strategies \cite{glymour2019,vowels2023}.

Constraint-based methods recover the graph by testing conditional independence (CI) relations. The PC algorithm \cite{spirtes2000} starts from a complete undirected graph, removes edges between conditionally independent pairs, orients v-structures, and applies Meek's rules \cite{meek1995} to produce a Completed Partially Directed Acyclic Graph (CPDAG); the order-independent variant of Colombo and Maathuis \cite{colombo2014} is commonly adopted. The Fast Causal Inference (FCI) algorithm \cite{spirtes1995}, \cite{ZHANG20081873} relaxes PC's assumption of causal sufficiency, allowing for latent confounders and outputting a Partial Ancestral Graph (PAG) with additional edge marks that signal unmeasured common causes.

Score-based methods formulate structure learning as optimization. The Greedy Equivalence Search (GES) \cite{chickering2002} searches the space of equivalence classes by greedily adding and then removing edges to maximize the Bayesian Information Criterion (BIC), with provable consistency in the large-sample limit.

Functional approaches exploit distributional assumptions for full identifiability. The Linear Non-Gaussian Acyclic Model (LiNGAM) \cite{shimizu2006} leverages non-Gaussianity of disturbance terms in the structural equation model $X_i = \sum_j b_{ij} X_j + e_i$ to identify the complete causal DAG, not just its equivalence class. The DirectLiNGAM variant \cite{shimizu2011} iteratively identifies root-cause variables via independence measures and regresses out their effects until a full causal ordering is obtained.

Continuous optimization methods recast the combinatorial DAG search as a differentiable problem. NOTEARS \cite{zheng2018} introduced a smooth acyclicity constraint via the matrix exponential, $h(\mathbf{W}) = \operatorname{tr}(e^{\mathbf{W} \circ \mathbf{W}}) - d = 0$, enabling gradient-based optimization. GOLEM \cite{ng2020} replaces the least-squares objective with a Gaussian log-likelihood and treats acyclicity as a soft penalty, while DAGMA \cite{bello2022} proposes a log-determinant characterization with better-behaved gradients and faster convergence.

\subsection{Markov Blanket}
\label{subsec:markov-blanket}

The Markov blanket of a variable provides a principled answer to the question: which variables are sufficient---and necessary---to predict a given target? Formally introduced by Pearl \cite{pearl1988}, the Markov blanket of a target variable $T$ in a joint distribution $P(V)$ is the minimal subset $\mathrm{MB}(T) \subset V \setminus \{T\}$ such that $T$ is conditionally independent of all remaining variables given $\mathrm{MB}(T)$:

\begin{equation}
\label{eq:mb-definition}
T \perp\!\!\!\perp \bigl(V \setminus (\mathrm{MB}(T) \cup \{T\})\bigr)
  \;\Big|\; \mathrm{MB}(T).
\end{equation}

When the distribution is faithful to a DAG $\mathcal{G}$, the Markov blanket has a well-known graphical characterization: $\mathrm{MB}(T)$ consists of the parents, children, and co-parents (also called spouses, i.e., other parents of $T$'s children) of $T$ in $\mathcal{G}$ \cite{pearl1988}. Parents represent direct causes of $T$, children represent its direct effects, and co-parents are variables that share a common effect with $T$ and therefore become informative through explaining-away reasoning.

The Markov blanket defines the theoretically optimal feature set for predicting $T$: no variable outside $\mathrm{MB}(T)$ carries additional predictive information, and removing any member would result in information loss \cite{koller1996,tsamardinos2003relevancy}. Unlike wrapper or filter methods that rank features by statistical association, the Markov blanket provides a causally grounded selection: the retained features are precisely those that participate in the causal mechanisms immediately surrounding the target \cite{pellet2008,yu2020survey}.

\subsection{Statistical Tests and Validation Metrics}
To support the interpretation of clustering results, a set of complementary statistical tests and validation metrics was employed, each capturing different aspects of clustering behavior. Partition similarity between clustering solutions can be evaluated using the Adjusted Rand Index (ARI) and Adjusted Mutual Information (AMI). These metrics quantify the agreement between two partitions while correcting for chance, with values closer to 1 indicating stronger similarity.

The relationship between clustering solutions can also be analyzed using the chi-square test of independence applied to the contingency table of cluster assignments. This test evaluates whether two partitions are statistically associated. The strength of this association can be quantified using Cramér’s V, which ranges from 0 (no association) to 1 (perfect association).

To assess differences in educational performance across clusters, one-way ANOVA can be used to test for differences in mean Saeb scores. Effect sizes may be computed using eta-squared ($\eta^2$) and omega-squared ($\omega^2$), providing measures of the proportion of variance explained by cluster membership. Homogeneity of variances can be evaluated using Levene’s test.

Given the potential violation of parametric assumptions, the Kruskal--Wallis test may also be applied as a non-parametric alternative, assessing whether score distributions differ across clusters. Finally, the internal quality of clustering solutions can be evaluated using the silhouette score and the Davies--Bouldin index. The silhouette score measures the degree of separation between clusters, with higher values indicating better-defined groupings, while the Davies--Bouldin index evaluates cluster compactness and separation, with lower values indicating better clustering structure. Together, these metrics provide a comprehensive and multi-perspective evaluation of clustering performance, combining structural, statistical, and geometric criteria.

\section{Methodology}
This study adopts a comparative methodological framework designed to evaluate how the selection of variables---particularly those derived from causal discovery---affects the identification and interpretation of school typologies. The methodological procedures adopted in this study are described below.

\subsection{Data Preprocessing}
Given the heterogeneity of the variables, a normalization step was applied prior to clustering. Continuous variables were scaled to the $[0,1]$ interval using Min--Max normalization, ensuring comparability across dimensions and preventing scale dominance in distance-based methods. Binary variables were preserved in their original form. This precaution is essential in distance-based methods, as it prevents variables with larger numerical ranges from exerting a disproportionate influence on the formation of clusters.

\subsection{Casual Features}
To identify the causal features associated with the target variable, multiple causal discovery algorithms were applied, including DAGMA, FCI, GES, and DirectLiNGAM. After estimating the candidate DAGs, we extracted the Markov blanket of the target variable, which provides the minimal set of variables that carry causal information about the outcome. This subset was then used as the causal feature set for the clustering analysis. The resulting graphs were compared in terms of structural plausibility and consistency with established knowledge in the educational domain. 

\subsection{Hierarchical Clustering}
The study applies hierarchical agglomerative clustering using Euclidean distance and Ward’s linkage to form compact and interpretable groups. Eight clusters were selected based on dendrogram inspection. Two analytical scenarios were considered: a baseline scenario, which uses all preprocessed variables to represent the traditional descriptive approach, and a causal scenario, which uses only the variables belonging to the Markov blanket of the target variable. This design allows assessing whether causal feature selection produces more coherent and performance‑relevant school typologies.

\subsection{Evaluation Framework}
The comparison between baseline and causal scenarios was conducted from three complementary perspectives. Structural differences between clustering solutions were assessed using partition similarity metrics and association measures, namely ARI, AMI, chi-square, and Cramér’s V. The impact of clustering on educational performance was evaluated using ANOVA, Levene’s test, effect size measures ($\eta^2$ and $\omega^2$), and the Kruskal--Wallis test. The internal geometric quality of clustering solutions was examined using silhouette score and Davies--Bouldin index.

This multi-criteria evaluation framework allows the comparison of clustering solutions not only in terms of structural similarity, but also with respect to performance differentiation and geometric consistency.

\section{Results and Discussion}
This section presents the results of the comparative analysis between the baseline and causal clustering scenarios. The findings are organized into three complementary dimensions: (i) structural differences between clustering solutions, (ii) impact of clustering on educational performance, and (iii) internal geometric quality of the clusters.

\subsection{Selection of the Causal Graph}
\begin{figure}[htbp]
	\centering
	\includegraphics[width=0.7\linewidth]{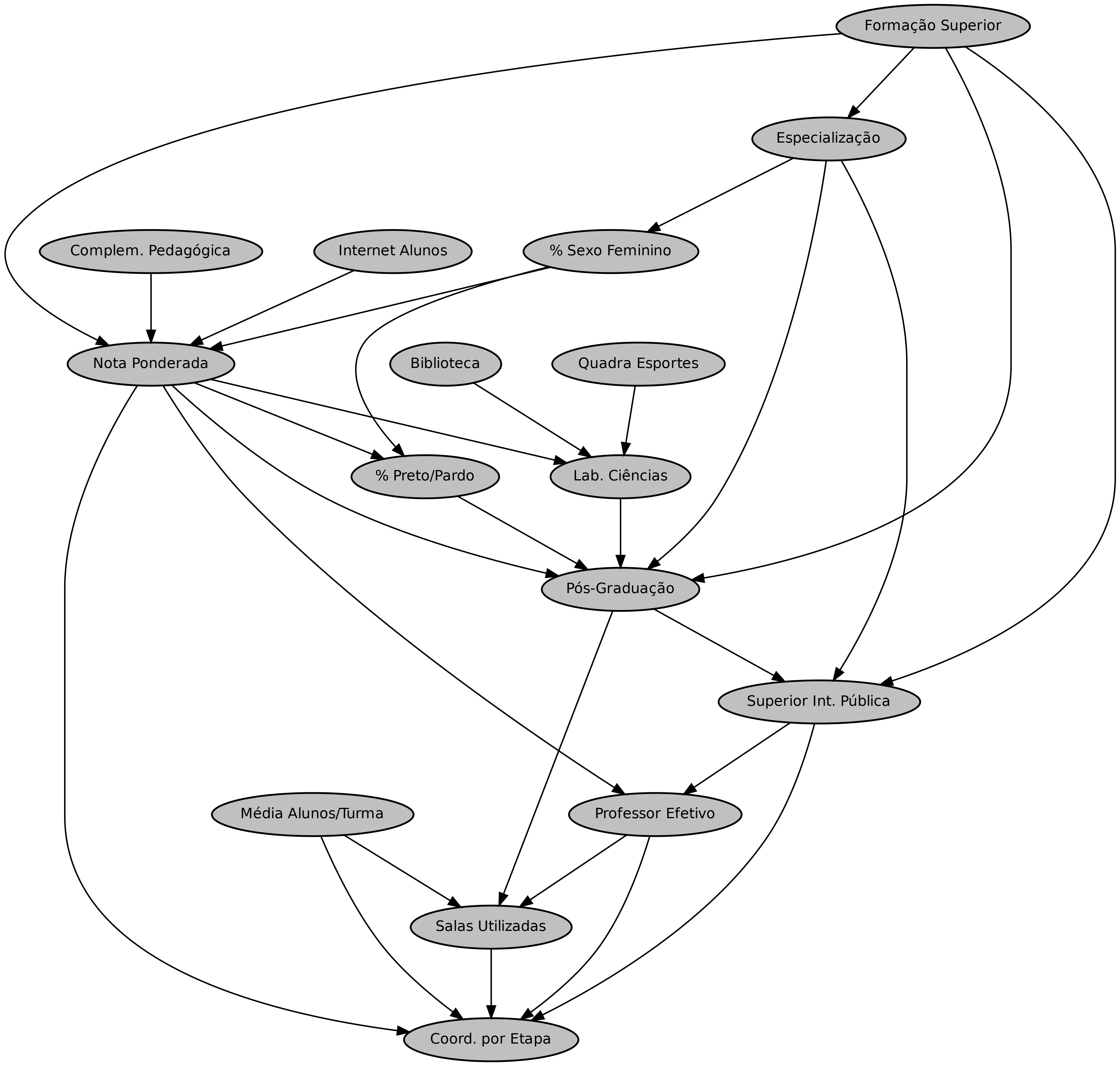}
	\caption{Causal diagram generated by the DirectLiNGAM algorithm.}
	\label{fig:censofulldirectlingam}
\end{figure}

Among the candidate causal models, the DAG obtained through the DirectLiNGAM algorithm exhibited the most coherent structure, capturing relationships consistent with theoretical and empirical expectations in education. Therefore, this DAG was selected as the basis for extracting the Markov blanket of the target variable. The selected causal graph is presented in Figure~\ref{fig:censofulldirectlingam}.

\subsection{Structural Differences Between Clustering Solutions}
\begin{figure}[htbp]
\centering
\begin{subfigure}{0.48\textwidth}
\centering
\includegraphics[width=\linewidth]{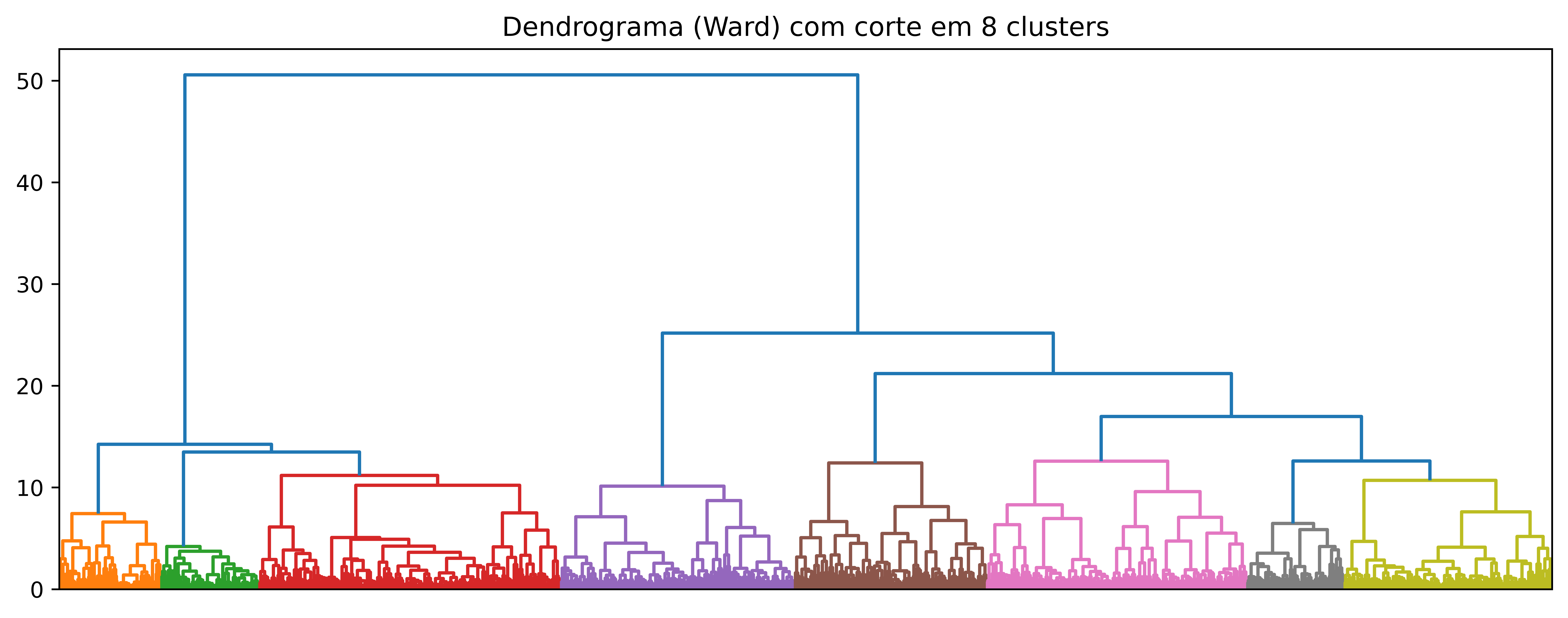}
\caption{Baseline}
\end{subfigure}
\hfill
\begin{subfigure}{0.48\textwidth}
\centering
\includegraphics[width=\linewidth]{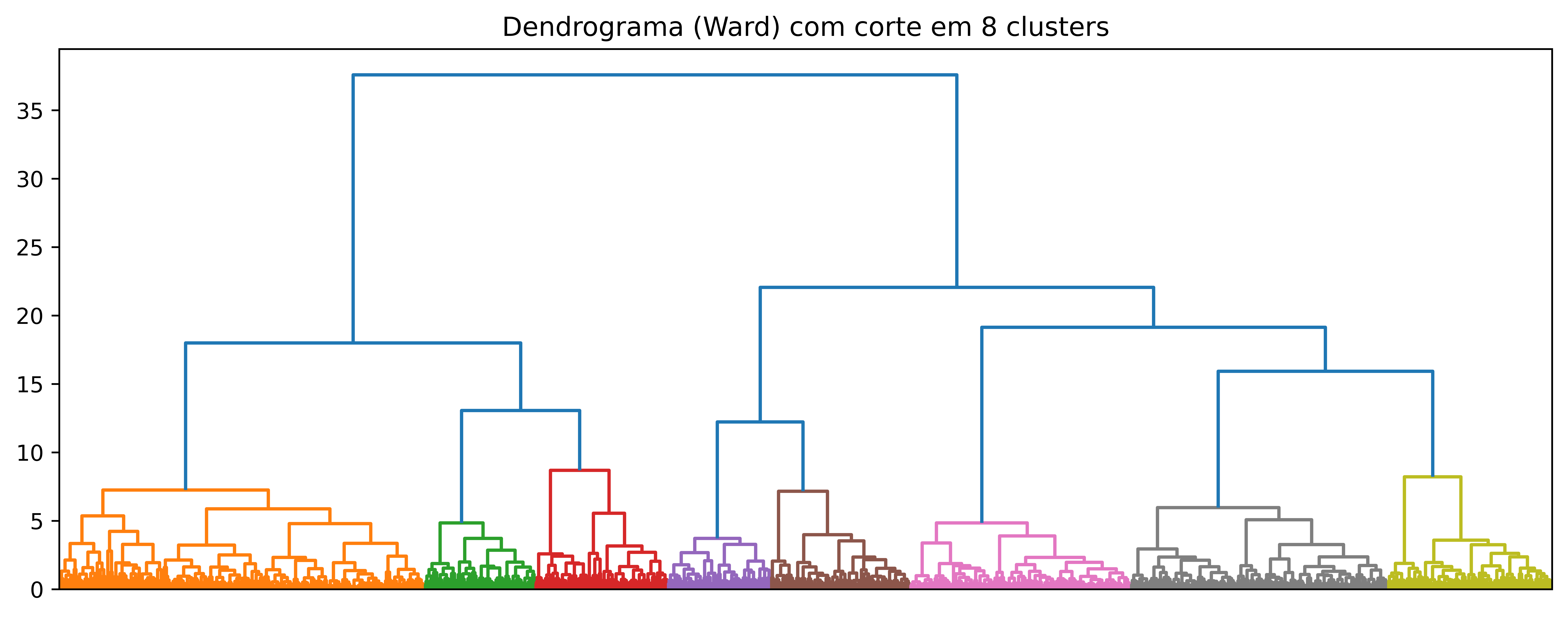}
\caption{Causal}
\end{subfigure}
\caption{Hierarchical clustering structures under baseline and causal scenarios}
\label{fig:dendrogram_comparison}
\end{figure}

The comparison between the baseline and causal clustering solutions reveals a moderate level of structural agreement combined with substantial reorganization of observations. The hierarchical structure of both clustering solutions is illustrated in Figure~\ref{fig:dendrogram_comparison}, highlighting differences in cluster formation between the baseline and causal scenarios. As listed in Table~\ref{tab:similarity_metrics}, the partition similarity metrics indicate partial alignment between the two configurations, with an Adjusted Rand Index (ARI) of 0.338 and an Adjusted Mutual Information (AMI) of 0.470. These values suggest that, although some structural patterns are preserved, the cluster assignments differ significantly across scenarios.

\begin{table}[htbp]
\centering
\caption{Similarity Between Clustering Solutions}
\label{tab:similarity_metrics}
\small
\begin{tabular}{lcc}
\toprule
Metric & Value & Interpretation \\
\midrule
Adjusted Rand Index (ARI) & 0.338 & Moderate agreement \\
Adjusted Mutual Information (AMI) & 0.470 & Moderate agreement \\
Cramér's V & 0.539 & Moderate-to-strong association \\
Chi-square ($\chi^2$) & 3691.79 & Strong dependence ($p < 0.001$) \\
\bottomrule
\end{tabular}
\end{table}

This result is reinforced by the chi-square test of independence, which strongly rejects the null hypothesis ($\chi^{2} = 3691.79$, $p < 0.001$), indicating that the two partitions are not independent. The strength of this association, measured by Cramér’s V (0.539), points to a moderate-to-strong relationship between the clustering structures.

\begin{figure}[htbp]
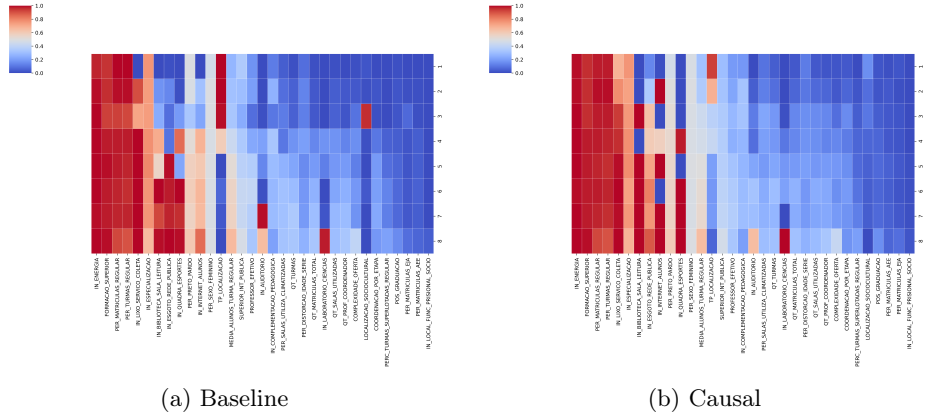

\centering
\begin{subfigure}{0.48\textwidth}
\centering
\includegraphics[width=\linewidth]{baseline_heatmap.png}
\caption{Baseline}
\end{subfigure}
\hfill
\begin{subfigure}{0.48\textwidth}
\centering
\includegraphics[width=\linewidth]{causal_heatmap_full.png}
\caption{Causal}
\end{subfigure}
\caption{Standardized variable profiles by cluster}
\label{fig:heatmap_comparison}
\end{figure}

The contingency analysis further indicates that certain clusters remain relatively stable, while others undergo substantial redistribution, particularly among intermediate structural profiles. This suggests that causal feature selection does not merely replicate the baseline clustering but instead reorganizes the data in a meaningful way, refining distinctions that are less evident when all variables are considered simultaneously. The cluster profiles based on standardized variables are shown in Figure~\ref{fig:heatmap_comparison}, allowing a visual comparison of structural patterns between the two scenarios.

\subsection{Impact of Clustering on Educational Performance}

As listed in Table~\ref{tab:performance_tests}, the relationship between cluster membership and educational performance was evaluated using both parametric and non-parametric statistical tests.

\begin{table}[htbp]
\centering
\caption{Statistical Comparison of Performance Across Clusters}
\label{tab:performance_tests}
\small
\begin{tabular}{lcccc}
\toprule
Scenario & ANOVA (F) & $\eta^2$ & $\omega^2$ & Kruskal--Wallis \\
\midrule
Baseline & 10.29 & 0.085 & 0.077 & 67.48 \\
Causal & 16.97 & 0.133 & 0.125 & 109.76 \\
\bottomrule
\end{tabular}
\end{table}

In both scenarios, the ANOVA results indicate statistically significant differences in the weighted Saeb score across clusters. However, the magnitude of these differences differs substantially between configurations.

\begin{figure}[htbp]
\centering
\begin{subfigure}{0.48\textwidth}
\centering
\includegraphics[width=\linewidth]{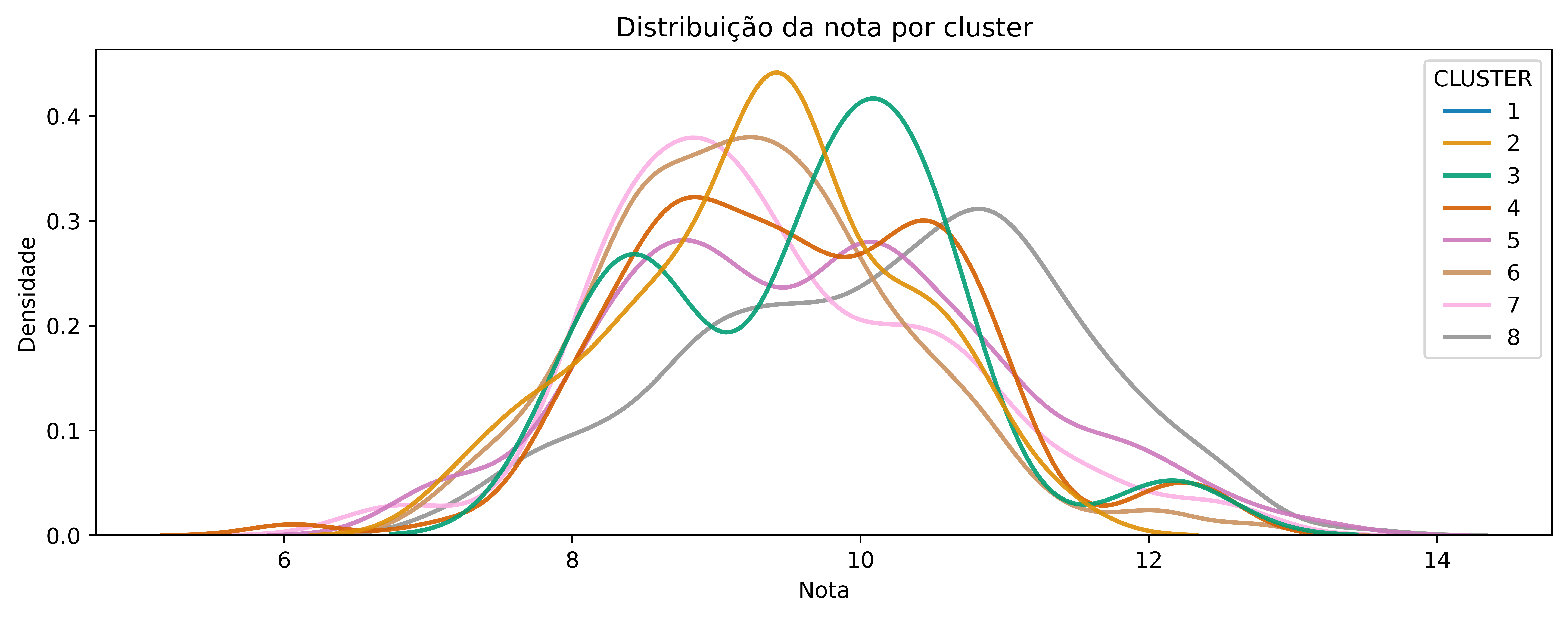}
\caption{Baseline}
\end{subfigure}
\hfill
\begin{subfigure}{0.48\textwidth}
\centering
\includegraphics[width=\linewidth]{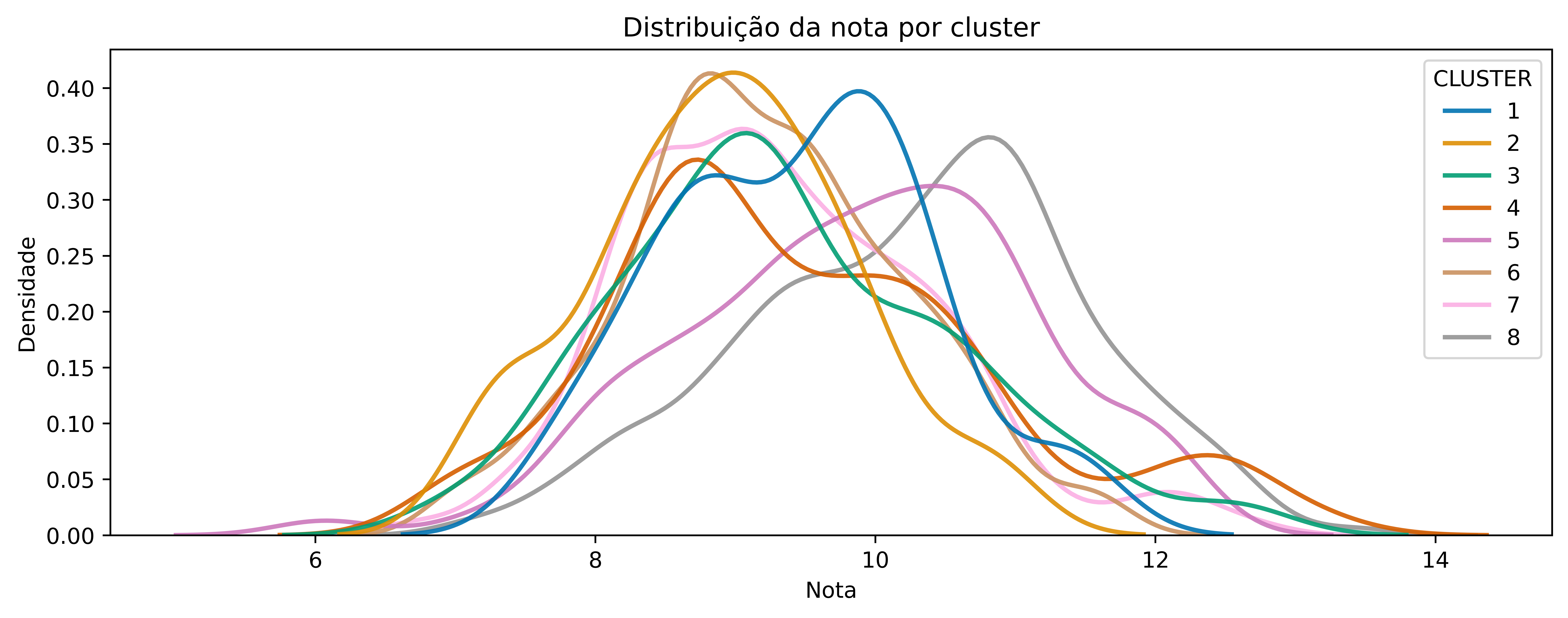}
\caption{Causal}
\end{subfigure}
\caption{Distribution of weighted Saeb scores across clusters}
\label{fig:kde_comparison}
\end{figure}

In the baseline scenario, the ANOVA statistic is $F = 10.29$ ($p < 0.001$), with effect sizes of $\eta^2 = 0.085$ and $\omega^2 = 0.077$, indicating a moderate explanatory power. Additionally, the Levene test rejects the assumption of homogeneity of variances ($p = 0.003$), suggesting instability in the distribution of scores across clusters.

In contrast, the causal scenario exhibits stronger statistical separation, with $F = 16.97$ ($p < 0.001$), and increased effect sizes ($\eta^2 = 0.133$ and $\omega^2 = 0.125$). The Levene test ($p = 0.052$) indicates that the assumption of homoscedasticity is no longer violated, suggesting a more stable distribution across clusters.

The Kruskal--Wallis test confirms these findings, with the statistic increasing from 67.48 in the baseline scenario to 109.76 in the causal configuration, reinforcing the robustness of the observed differences under weaker distributional assumptions.

These results are consistent with the distribution plots presented in Figure~\ref{fig:kde_comparison}, where the causal clustering exhibits clearer separation between score distributions, while the baseline scenario shows substantial overlap among clusters.

Overall, the results indicate that clustering based on causal features improves the discriminative capacity of the typologies with respect to educational performance.

\subsection{Internal Geometric Quality of Clusters}
As summarized in Table~\ref{tab:internal_validation}, the internal quality of the clustering solutions was assessed using silhouette score and Davies--Bouldin index. The baseline clustering exhibits a low silhouette score and a high Davies--Bouldin index, indicating weak separation between clusters and high internal dispersion. This suggests that, although clusters may present distinguishable average profiles, they are not well separated at the instance level in the feature space.

\begin{table}[htbp]
\centering
\caption{Internal Clustering Quality Metrics}
\label{tab:internal_validation}
\small
\begin{tabular}{lcc}
\toprule
Scenario & Silhouette Score & Davies--Bouldin Index \\
\midrule
Baseline & 0.1211 & 2.0120 \\
Causal & 0.3496 & 1.2040 \\
\bottomrule
\end{tabular}
\end{table}

In contrast, causal clustering indicates a substantial improvement in both metrics. The silhouette score increases from 0.1211 to 0.3496, indicating stronger cohesion and separation, while the Davies--Bouldin index decreases from 2.0120 to 1.2040, reflecting more compact and well-defined clusters. This improvement suggests that the feature space defined by the Markov blanket leads to a clearer geometric structure, in which schools are grouped into more coherent and distinct typologies.

The heatmap reveals a seamless transition between clusters. As we move down the rows, the performance indicators shift gradually rather than showing sharp boundaries, suggesting that the school typologies exist along a continuous spectrum. This partially explains the discrepancy between visual inspection of aggregated profiles and geometric evaluation metrics. Overall, the results indicate that causal feature selection not only improves the explanatory relationship between clusters and performance but also enhances the internal consistency and separability of the clustering structure.

\subsection{Discussion}
The comparison between baseline and causal clustering shows that the inclusion of causal features does not merely generate an alternative partition but produces a more structured reorganization of the data. Although similarity metrics indicate only moderate agreement between the two solutions, the redistribution of observations suggests that causal features refine the latent structure of the dataset. Clusters that appeared diffuse or partially overlapping in the baseline scenario become more clearly delineated when causal variables are used, indicating that the causal configuration isolates dimensions more closely tied to the mechanisms shaping school characteristics. This reorganization should therefore be interpreted not as instability but as a shift from descriptive grouping toward a more structurally meaningful partition.

The results also reveal that clusters derived from causal features exhibit stronger and more consistent associations with educational performance. Effect sizes and test statistics increase substantially in the causal scenario, indicating that cluster membership becomes more informative with respect to Saeb scores. In practical terms, schools grouped under the causal configuration are more homogeneous in performance and more distinct across clusters. This has important implications for educational analysis: while traditional clustering may identify broad structural typologies, it does not necessarily capture patterns that matter for performance evaluation. By contrast, the causal approach yields groupings that align more closely with outcome variation, making them more suitable for policy analysis and targeted interventions. The improved homogeneity of variances further suggests that causal clusters represent more stable populations, reducing the risk of misleading comparisons between heterogeneous groups.

\section{Conclusion}
This study demonstrates that integrating causal reasoning into feature selection can substantially enhance the interpretability and analytical value of clustering results. Traditional pipelines often rely on statistical association or geometric criteria, which may introduce redundancy and obscure meaningful structures. The proposed approach, by contrast, incorporates domain knowledge and causal structure into the selection process, producing clusters that are not only geometrically coherent but also more relevant for explanatory and decision‑support purposes. The multi‑criteria evaluation framework adopted here—combining structural similarity, performance association, and geometric validation—provides a more comprehensive understanding of clustering behavior than any single metric alone. These findings suggest that causal feature selection is a promising direction for unsupervised learning in applied domains where interpretability and outcome relevance are essential. Despite these contributions, the study has limitations. The identification of causal features depends on domain assumptions and prior knowledge, which may introduce subjectivity. The analysis is also restricted to a single dataset and educational context, limiting generalizability. Future work will concentrate on expanding the approach to a nationwide analysis, allowing for broader generalization of the findings.

\bibliographystyle{splncs04}
\bibliography{referencias}

\end{document}